\documentstyle[11pt]{article}
\begin{document}
\title{ Asymptotic Behaviour of Inhomogeneous  String Cosmologies}
\author{Kerstin E. Kunze\\
\\
{\normalsize\sl Institut f\"{u}r Theoretische Physik, ETH, 
CH-8093 Z\"{u}rich, Switzerland}}
\date{}
\maketitle

\begin{abstract}
The asymptotic behaviour at late times 
of inhomogeneous axion-dilaton cosmologies
is investigated. 
The space-times considered here 
admit two abelian space-like Killing vectors.
These space-times evolve towards an anisotropic universe 
containing gravitational radiation.
Furthermore, a peeling-off behaviour of the Weyl tensor and
the antisymmetric tensor field strength is found.
The relation to the pre-big-bang scenario
is briefly discussed.
\end{abstract}

\section{Introduction}
The low energy effective action of string theory provides 
cosmological models that might be applicable just below the string scale 
in the very early universe. In the pre-big-bang scenario the universe 
starts in a low curvature, low coupling regime and then enters a stage
of dilaton driven kinetic inflation \cite{gasp}.
This picture has been further developed in \cite{buon}. 
Initially the universe consists of a bath of gravitational and
dilatonic waves. Some of these collapse leading to the birth of
a baby inflationary universe.

The aim here is to investigate the asymptotic future
states of a large class of inhomogeneous string cosmologies.
Late times in the usual general relativistic setting correspond to
early times in the pre-big-bang picture.
As will be seen, this provides further evidence that 
the Milne model is not 
a generic initial state for the pre-big-bang scenario.
The issue of a generic initial state has also been investigated 
in \cite{clan} with the emphasis on orthogonal and tilted Bianchi
string cosmologies. There it was found that a plane wave background is
a likely attractor and not the Milne universe as was conjectured
in \cite{Buon2}. Einstein-Rosen string cosmologies were discussed in 
\cite{jk1} \cite{p3}.

In general relativity the asymptotic evolution of 
gravitational radiation from a bounded source was investigated 
\cite{sachs}  and it was found that the Riemann tensor
shows a ``peeling off'' behaviour, meaning that different components
of the Weyl tensor evolve as different powers of the radial
coordinate.

A cosmological peeling off theorem was formulated by Carmeli and Feinstein
\cite{carm} by using earlier results of Stachel \cite{sta} in the
case of cylindrical gravitational waves.
Cosmological space-times of the Einstein-Rosen type can be obtained from
these cylindrical gravitational wave space-times by interchanging the
non-ignorable coordinates. 
The cosmological peeling off theorem decribes the asymptotic future 
behaviour  of inhomogeneous
vacuum space-times admitting two abelian Killing vectors.
The general form of the metric allows space-time homogeneity to be
broken along one spatial direction, so in general solutions will depend on
one spacelike and one timelike coordinate. However, this type of metric
also encompasses the --  by definition -- spatially homogeneous 
metrics of Bianchi type I-VII \cite{tom}\cite{carm}.
In \cite{carm} it was shown that the Weyl tensor shows a peeling off
behaviour in terms of the timelike variable. At late times
the dominant contribution comes from the Weyl scalar $\Psi_{4}$ which
indicates that the space-time approaches at late times a plane wave
space-time.

A similar peeling off behaviour will be found in the case of
axion-dilaton string cosmology.

In the bosonic sector of the low energy effective action of a string
theory general relativity coupled to two massless scalar fields is
obtained in the so-called Einstein frame.

In four dimensions the action is given in the string frame by 
\cite{r4}
\begin{equation}
S=\int d^4 x\sqrt{-g}e^{-\phi }(R+g^{\alpha \beta }\partial _\alpha \phi
\partial _\beta \phi -\frac 1{12}H^{\alpha \beta \gamma }H_{\alpha \beta
\gamma })
\end{equation}
where the antisymmetric tensor field strength $H_{\alpha \beta \gamma
}=\partial _{[\alpha }B_{\beta \gamma ]} \label{H}$ is introduced.
 
Applying the conformal transformation
\begin{equation}
g_{\alpha \beta }\rightarrow e^{-\phi }g_{\alpha \beta }.
\end{equation}
this can be written in the usual Einstein-Hilbert form.

The physical frame is the string frame since in this frame strings move
along geodesics. However, since the Weyl tensor is conformally invariant
it is justified to use the Einstein frame to find  solutions and 
take advantage of the formalism developed for general relativity.

In the Einstein frame the equations of motion are given by

\cite{r4} 
 
\begin{eqnarray}
{R}_{\mu\nu}-\frac{1}{2}{g}_{\mu\nu}{R}
&=&
^{(\phi)}{T}_{\mu\nu}+ ^{(H)}{T}_{\mu\nu}
\\
{\nabla}_{\mu}\left[\exp(-2\phi){H}^{\mu\nu\lambda}
\right]&=&0\label{H-eq}\\
{\large{\Box}}\phi+\frac{1}{6}e^{-2\phi}{H}_{\alpha\beta\gamma}
{H}^{\alpha\beta\gamma}
&=&0
\end{eqnarray}

where 
\begin{eqnarray}
^{(\phi)}{T}_{\mu\nu}&=&
\frac{1}{2}(\phi_{,\mu}\phi_{,\nu}
-\frac{1}{2}{g}_{\mu\nu}{g}^{\alpha\beta}
\phi_{,\alpha}\phi_{,\beta})\label{tphi}\\
^{(H)}{T}_{\mu\nu}&=&
\frac{1}{12}e^{-2\phi}\left(
3{H}_{\mu\lambda\kappa}{H}_{\nu}^{\;\;\lambda\kappa}
-\frac{1}{2}{g}_{\mu\nu}{H}_{\alpha\beta\gamma}
{H}^{\alpha\beta\gamma}\right)\\
\label{tH}
\end{eqnarray}

Furthermore in four dimensions the antisymmetric tensor field strength can
be written in terms of the scalar field, $b$, as follows
\begin{eqnarray}
H^{\mu\nu\lambda}=e^{2\phi}\epsilon^{\rho\mu\nu\lambda}
b_{,\rho}.\label{H1}
\end{eqnarray}

\section{Equations of motion for axion and dilaton} 

Metrics admitting two spacelike abelian Killing vectors can be written in
the following general form \cite{grif}

\begin{eqnarray}
ds^{2}=2e^{-M}dudv-\frac{2e^{-U}}{Z+\bar{Z}}
\left(dx+iZdy\right)
\left(dx-i\bar{Z}dy\right)
\label{1}
\end{eqnarray}

$M$ and $U$ are real functions depending on 
the null variables $u$ and $v$.
$Z$ is a complex function depending on 
$u$ and $v$. 
Complex conjugation is denoted by a bar.

It is convenient to introduce a null tetrad with 
two real null
vectors, ${\bf l}$, ${\bf n}$, and two complex null vectors 
which are conjugates of one another, ${\bf m}$, 
${\bf \bar{m}}$.

They satisfy the following relations \cite{grif}\cite{chandra}
\begin{eqnarray}
l_{\mu}n^{\mu}=1 & m_{\mu}\bar{m}^{\mu}=-1
\end{eqnarray}

and the completeness relation
\begin{eqnarray}
g_{\mu\nu}=l_{\mu}n_{\nu}+n_{\mu}l_{\nu}
-m_{\mu}\bar{m}_{\nu}-\bar{m}_{\mu}m_{\nu}.
\end{eqnarray}

Useful formulae for quantities in the Newman-Penrose formalism are given
in the appendix.

Including the axion $b$ and dilaton $\phi$ and using Einstein's equations
the components of the Ricci tensor can be determined.
The energy momentum tensor is
given by
\begin{eqnarray}
T_{\mu\nu}=\frac{1}{2}[\phi_{,\mu}\phi_{,\nu}+
e^{2\phi}b_{,\mu}b_{,\nu}
-\frac{1}{2}g_{\mu\nu}
g^{\alpha\beta}
(\phi_{,\alpha}\phi_{,\beta}+e^{2\phi}b_{,\alpha}b_{,\beta})].
\end{eqnarray}

The following equations of motion are obtained

\begin{eqnarray}
2U_{uu}-(U_{u})^{2}+2M_{u}U_{u}
-4\frac{Z_{u}\bar{Z}_{u}}{(Z+\bar{Z})^{2}}&=&\phi_{u}^{2}
+e^{2\phi}b_{u}^{2}
\label{e6}\\
2U_{vv}-(U_{v})^{2}+2M_{v}U_{v}
-4\frac{Z_{v}\bar{Z}_{v}}{(Z+\bar{Z})^{2}}&=&\phi_{v}^{2}
+e^{2\phi}b_{v}^{2}
\label{e7}\\
2Z_{uv}-U_{u}Z_{v}-U_{v}Z_{u}-4\frac{Z_{u}Z_{v}}{Z+\bar{Z}}&=&0
\label{e8}\\
2M_{uv}+U_{u}U_{v}
-2\frac{Z_{u}\bar{Z}_{v}+\bar{Z}_{u}Z_{v}}{(Z+\bar{Z})^{2}}
&=&\phi_{u}\phi_{v}
+e^{2\phi}b_{u}b_{v}
\label{e9}\\
U_{uv}&=&U_{u}U_{v}
\label{e10}\\
2\phi_{uv}-U_{u}\phi_{v}-U_{v}\phi_{u}-2e^{2\phi}b_{u}b_{v}&=&0
\label{e11}\\
2b_{uv}-U_{u}b_{v}-U_{v}b_{u}+2\phi_{u}b_{v}+2\phi_{v}b_{u}&=&0
\label{e11a}
\end{eqnarray}

Equation (\ref{e9}) is an integrability condition for
equations (\ref{e6}) and (\ref{e7}).
The general solution of equation (\ref{e10}) is given by
\cite{grif}

\begin{eqnarray}
e^{-U}=f(u)+g(v)
\end{eqnarray}

where $f$ and $g$ are arbitrary functions of their arguments.
This can be used to write the equations of motion in a more compact
way by using $f$ and $g$ as coordinates.

The equations can be simplified further by introducing a new
function $S$, defined as follows \cite{grif}
\begin{eqnarray}
e^{-M}=\frac{f'g'}{\sqrt{f+g}}e^{-S}
\label{e-M}
\end{eqnarray}

where the prime denotes a derivative w.r.t. to $u$ ($v$)
for $f$ ($g$). 

So this reduces the system of equations (\ref{e6})--(\ref{e11a})
to the following

\begin{eqnarray}
2Z_{fg}+\frac{1}{f+g}\left(Z_{f}+Z_{g}\right)
-4\frac{Z_{f}Z_{g}}{Z+\bar{Z}}&=&0
\label{ernst}\\
2\phi_{fg}+\frac{1}{f+g}\left(\phi_{f}+\phi_{g}\right)
-2e^{2\phi}b_{f}b_{g}
&=&0\\
2b_{fg}+\frac{1}{f+g}\left(b_{f}+b_{g}\right)
+2\phi_{f}b_{g}+2\phi_{g}b_{f}&=&0\label{b_fg}
\end{eqnarray}

\begin{eqnarray}
S_{f}&=&-2(f+g)\left(\frac{Z_{f}\bar{Z}_{f}}{(Z+\bar{Z})^{2}}
+\frac{1}{4}(\phi_{f}^{2}+e^{2\phi}b_{f}^{2})\right)
\label{s-f}\\
S_{g}&=&-2(f+g)\left(\frac{Z_{g}\bar{Z}_{g}}{(Z+\bar{Z})^{2}}
+\frac{1}{4}(\phi_{g}^{2}+e^{2\phi}b_{g}^{2})\right).
\label{s-g}
\end{eqnarray}

Therefore the components of the Weyl tensor are given by

\begin{eqnarray}
\Psi_{0}&=&\frac{e^{M}}{(Z+\bar{Z})^{2}}(g')^{2}
\left[(Z+\bar{Z})\left(\bar{Z}_{gg}+\left[\frac{3}{2}\frac{1}{f+g}
-2(f+g)\left[\frac{Z_{g}\bar{Z}_{g}}{(Z+\bar{Z})^{2}}+
\frac{1}{4}(\phi_{g}^{2}+e^{2\phi}b_{g}^{2})\right]\right]\bar{Z}_{g}\right)
\right.\nonumber\\
& &\;\;\;\;\;\;\;\;\;\;\;\;\;\;\;\;\;\;\;\;\;\;\;\;\;
-2(\bar{Z}_{g})^{2}
\left. \right]\\
\Psi_{2}&=&-\frac{e^{M}}{12}f'g'
\left[\frac{3}{(f+g)^{2}}-12\frac{Z_{f}\bar{Z}_{g}}
{(Z+\bar{Z})^{2}}-(\phi_{f}\phi_{g}+e^{2\phi}b_{f}b_{g})\right]\\
\Psi_{4}&=&\frac{e^{M}}{(Z+\bar{Z})^{2}}(f')^{2}
\left[(Z+\bar{Z})\left(Z_{ff}+\left[\frac{3}{2}\frac{1}{f+g}
-2(f+g)\left[\frac{Z_{f}\bar{Z}_{f}}{(Z+\bar{Z})^{2}}+
\frac{1}{4}(\phi_{f}^{2}+e^{2\phi}b_{f}^{2})\right]\right]Z_{f}\right)
\right.\nonumber\\
& &\;\;\;\;\;\;\;\;\;\;\;\;\;\;\;\;\;\;\;\;\;\;\;\;\;    
-2(Z_{f})^{2}
\left. \right]
\end{eqnarray}

\section{Exact solutions}

In this section several exact  solutions are 
discussed. These are of interest since they will already show
some of the behaviour of the more general space-times 
which will be discussed in the next section.
Introducing two new variables,

\begin{eqnarray}
f=t-z\;\;\;\;\;\;\;\;\;\;\;\;\;\;
g=t+z
\end{eqnarray}

and the Ernst potential $Z$ may be conveniently written as

\begin{eqnarray}
Z=\zeta+i\omega
\end{eqnarray}

where $\zeta$ and $\omega$ are real functions.
When the imaginary part of the Ernst potential vanishes,
the two commuting Killing vectors become hypersurface
orthogonal. In this case the metric is globally
diagonalizable.

In the special case of a diagonal and homogeneous ansatz 
equation (\ref{ernst}) reads
\begin{eqnarray}
\zeta(\zeta_{tt}+t^{-1}\zeta_{t})-\zeta_{t}^{2}=0
\end{eqnarray}

which is solved by
\begin{eqnarray}
\zeta=at^{c}
\label{zeta}
\end{eqnarray}
with $a$ and $c$ constants.

The equations for the dilaton and axion are given by
\begin{eqnarray}
\phi_{tt}+t^{-1}\phi_{t}-e^{2\phi}b_{t}^{2}&=&0\\
b_{tt}+t^{-1}b_{t}+2\phi_{t}b_{t}&=&0
\end{eqnarray}

which is solved by \cite{bata} \cite{jk1}

\begin{eqnarray}
e^{\phi}&=&\cosh(N\xi)+\sqrt{1-\frac{B^{2}}{N^{2}}}\sinh(N\xi)
\label{phi}\\
b(\xi)&=&\frac{N}{B}\frac{\sinh(N\xi)+\sqrt{1-\frac{B^{2}}{N^{2}}}
\cosh(N\xi)}{\cosh(N\xi)+\sqrt{1-\frac{B^{2}}{N^{2}}}\sinh(N\xi)}
\label{b}
\end{eqnarray}

where $dt=td\xi$.
This solution can actually be obtained by an SL(2;$I\!\!R$) transformation
from a pure dilaton solution.

The components of the Weyl tensor are given by
 
\begin{eqnarray}
\Psi_{0}&=&\frac{e^{M}}{16}(g')^{2}c
\left[1-c^{2}-N^{2}\right]t^{-2}\label{wey-kas1}\\
\Psi_{2}&=&-\frac{e^{M}}{48}f'g'\left[3(1-c^{2})-N^{2}
\right]t^{-2}\\
\Psi_{4}&=&\frac{e^{M}}{16}(f')^{2}c\left[1-c^{2}-N^{2}
\right]t^{-2}\label{wey-kas3}
\end{eqnarray}

Using (\ref{zeta}) as ansatz for $\zeta$ and solving (\ref{ernst})
for $\omega$ results in an inhomogeneous
solution for $\omega$  \cite{sta}.
In this case (\ref{ernst}) yields the following two
equations
\begin{eqnarray}
\zeta(\zeta_{tt}+t^{-1}\zeta_{t})-(\zeta_{t}^{2}-\omega_{t}^{2})
-\omega_{z}^{2}&=&0\\
\zeta(\omega_{tt}+t^{-1}\omega_{t}-\omega_{zz})-2\omega_{t}
\zeta_{t}&=&0.
\end{eqnarray}

For the special value $c=\frac{1}{2}$ this has the
solution
\begin{eqnarray}
\zeta&=&at^{\frac{1}{2}}
\label{com1}\\
\omega&=&\Theta(\kappa) F(t-z)+(1-\Theta(\kappa))G(t+z)
\label{com2}
\end{eqnarray}

where $\Theta(\kappa)$ is the step function
($\Theta(\kappa)=0$ for $\kappa\leq 0$; $\Theta(\kappa)=1$ for $\kappa>0$)
and $\kappa$ is an arbitrary real parameter.
Note that $\omega$ can either include $F$ or $G$, but not 
both.

Another solution in this class could be easily generated by
applying a mirror transformation. Mirror symmetries in the
context of metrics with two abelian Killing vectors were 
discussed in \cite{mirror}. The effect of such a transformation in the
class of backgrounds considered here is to exchange 
the background metric and the axion-dilaton content.
Applying such a transformation to the solution above (\ref{phi}), (\ref{b}),
(\ref{com1}), (\ref{com2})
results in a solution with a homogeneous
dilaton and an inhomogeneous, oscillating axion similar to the
solution found in \cite{jk1}. Since the solution for the axion and
dilaton can be obtained by an SL(2;$I\!\!R$) transformation 
the space-time metric in the mirror image is not
genuinely diagonal.

The components of the Weyl tensor are given by

\begin{eqnarray}
\Psi_{0}&=&e^{M}(g')^{2}\left[
\frac{1}{32}(\frac{3}{4}-N^{2})t^{-2}+\frac{3}{8a^{2}}(1-\Theta(\kappa))^2
[G'(g)]^{2}t^{-1}\right.\nonumber\\
& &\left.
-\frac{i}{8a}(\frac{3}{4}-N^{2})(1-\Theta(\kappa))G'(g)t^{-\frac{3}{2}}
+\frac{i}{2a^{3}}(1-\Theta(\kappa))^3[G'(g)]^{3}t^{-\frac{1}{2}}
\right.\nonumber\\
& &\left.
-\frac{i}{2a}(1-\Theta(\kappa))G''(g)t^{-\frac{1}{2}}
\right]\\
\Psi_{2}&=&-\frac{e^{M}}{48}f'g'\left[
(\frac{9}{4}-N^{2})t^{-2}-\frac{12}{a^{2}}(1-\Theta(\kappa))\Theta(\kappa)
G'(g)F'(f)t^{-1}\right.\nonumber\\
& &\left.
-i\frac{3}{a}\left[\Theta(\kappa)F'(f)-(1-\Theta(\kappa))G'(g)\right]
t^{-\frac{3}{2}}\right]\\
\Psi_{4}&=&e^{M}(f')^{2}\left[
\frac{1}{32}(\frac{3}{4}-N^{2})t^{-2}+\frac{3}{8a^{2}}
\Theta(\kappa)^{2}[F'(f)]^{2}t^{-1}\right.\nonumber\\
& &\left.
+\frac{i}{8a}(\frac{3}{4}-N^{2})\Theta(\kappa)F'(f)t^{-\frac{3}{2}}
-\frac{i}{2a^{3}}\Theta(\kappa)^{3}[F'(f)]^{3}t^{-\frac{1}{2}}
\right.\nonumber\\
& &\left.
+\frac{i}{2a}\Theta(\kappa)F''(f)t^{-\frac{1}{2}}\right].
\end{eqnarray}

For this solution the equations for $S$, (\ref{s-f}), (\ref{s-g}),
have the form

\begin{eqnarray}
S_{f}&=&-\frac{1}{2}(\frac{1}{4}+N^{2})\frac{1}{f+g}-
\frac{1}{a^{2}}[\omega_{f}]^{2}\\
S_{g}&=&-\frac{1}{2}(\frac{1}{4}+N^{2})\frac{1}{f+g}-
\frac{1}{a^{2}}[\omega_{g}]^{2}.
\end{eqnarray}

This system of equations can be solved by introducing a new function
$\Omega(f,g)$ such that
\begin{eqnarray}
\Omega_{f}=[\omega_{f}]^{2}\;\;\;\;\;\;\;\;\;
\Omega_{g}=[\omega_{g}]^{2}\label{o2}
\end{eqnarray}

which implies, using that $\omega_{fg}=0$,
$$\Omega_{fg}=0.$$

This equation has the general solution
$$\Omega(f,g)=\Omega_{1}(f)+\Omega_{2}(g)$$

and hence using (\ref{o2})

\begin{eqnarray}
\Omega(f,g)=\int df(\omega_{f})^{2}+ \int dg (\omega_{g})^{2}.
\end{eqnarray}

Thus $S$ is given by

\begin{eqnarray}
S=-\frac{1}{2}(\frac{1}{4}+N^{2})\ln(f+g)-\frac{1}{a^{2}}\Omega+\ln k
\end{eqnarray}

where $k$ is an arbitrary constant.

With equation (\ref{e-M}) the factor $e^{M}$ in the expressions for
the components of the Weyl tensor is given by
\begin{eqnarray}
e^{M}=k(f'g')^{-1}t^{\frac{3}{8}-\frac{1}{2}N^{2}}
e^{-\frac{1}{a^{2}}\Omega}.
\end{eqnarray}

For $\Theta(\kappa)=1$ and $f=t-z$ fixed, then the components
of the Weyl tensor behave as $\Psi_{0}\sim e^{M}(g')^2t^{-2}$,
$\Psi_{2}\sim e^{M}f'g't^{-\frac{3}{2}}$, $\Psi_{4}\sim
e^{M}(f')^2 t^{-\frac{1}{2}}$ as $t\rightarrow\infty$.

A similar behaviour is found for $\Theta(\kappa)=0$ and $g=t+z$
fixed in the limit $t\rightarrow\infty$. In this case the behaviour 
of $\Psi_{0}$ and $\Psi_{4}$ is exchanged. In both cases a 
plane wave space-time is approached. However, the plane waves are in opposing
directions.

Furthermore, independent of $\Theta(\kappa)$, the pre-factor
$e^{M}$ is given by ${\cal O}(t^{\frac{3}{8}-\frac{1}{2}N^{2}})$

Finally, one more class of exact solutions will be discussed.
The metric is again assumed to be diagonal.
With $\zeta=e^{-V}$ the Ernst equation (\ref{ernst}) implies
\begin{eqnarray}
V_{tt}+\frac{1}{t}V_{t}-V_{zz}=0.
\end{eqnarray}

This is solved by \cite{grif} 
\begin{eqnarray}
V&=&-a\ln t+L_{1}\{A_{\omega}\cos[\omega(z+\alpha_{\omega})]J_{0}(\omega t)\}
+L_{2}\{B_{\omega}\cos[\omega(z+\beta_{\omega})]Y_{0}(\omega t)\}\nonumber\\
&&
+\sum_{i} d_{i}\cosh^{-1}\left(\frac{z+c_{i}}{t}\right)
\label{v-sol}
\end{eqnarray}

where $L_{1}\{ \}$ and $L_{2}\{ \}$ are arbitrary combinations of the
terms in curly brackets.
$J_{0}$ and $Y_{0}$ are Bessel functions of the first and second kind. 
$A_{\omega}$, $B_{\omega}$, $\alpha_{\omega}$ and $\beta_{\omega}$
are arbitrary constants.

The first term on its own describes a Kasner solution. The terms 
containing the Bessel functions will be investigated in the
next section. The last term introduces the so-called
gravitosolitonic contribution.

In the simplest case, the gravitosolitonic contribution
contains just one term and the constant $c$ is set to zero. The constant
pre-factor is $d$.

Neglecting the contributions from the Bessel functions
solutions by Wainwright et al. are recovered \cite{wain}.
These are related to the Kasner metrics by a soliton transformation
\cite{kitch} \cite{verd}.

Furthermore, for $d^{2}=a^{2}+3$ the metrics are the Ellis-MacCallum 
solutions of Bianchi type III, V and VI \cite{ellis} \cite{carm2}
\cite{verd}. Coordinates $\tilde{t}$, $\tilde{z}$  adapted to 
spatial homogeneity are given by \cite{carm2}\cite{verd}
$$t=e^{-2m\tilde{z}}\sinh 2m\tilde{t}\;\;\;\;\;\;\;\;\;
z=e^{-2m\tilde{z}}\cosh 2m\tilde{t}.$$
Depending on the parameter $a$ the models are of Bianchi type III,
V or VI.

As was pointed out by Carmeli et al. \cite{carm2}, in general
these space-times are not only singular at $\tilde{t}=0$ but
also at $\tilde{z}\rightarrow\pm\infty$. This behaviour depends
on the parameters $a$ and $d$. Only the Bianchi models
are singularity free at $\tilde{z}\rightarrow\pm\infty$.
If one is interested in a cosmological interpretation
this seems to confine solutions containing a gravitosolitonic
contribution to those which are spatially homogeneous.

Orthogonal Bianchi VI$_{h}$ models approach a plane wave space-time
at future timelike and null infinity. The other two approach
flat space-time \cite{hew} \cite{clan}.

\section{Asymptotic behaviour of the Weyl tensor and antisymmetric 
tensor field strength}

The Petrov type of the Weyl tensor describes the number of 
its distinct principal null directions.
It is determined by the roots of the following equation
\cite{chandra}
\begin{eqnarray}
\Psi_{4}b^{4}+4\Psi_{3}b^{3}+6\Psi_{2}b^{2}+4\Psi_{1}b
+\Psi_{0}=0
\end{eqnarray}

In the case of metrics of type (\ref{1}) the complex Weyl scalars 
$\Psi_{1}$ and $\Psi_{3}$ vanish. 
Therefore there are either four distinct or two double
roots.
So the space-time is of type I or of type D.

In order to find the Petrov type of a
space-time an algorithm developed in \cite{dinv} can be applied.
In the case of the metric considered here, the criterion for
the space-time being of type D is \cite{dinv} \cite{verd}
\begin{eqnarray}
I^{3}=27J^{2}
\end{eqnarray}

where $I\equiv\Psi_{0}\Psi_{4}+3\Psi_{2}^{2}$
and $J\equiv\Psi_{2}(\Psi_{0}\Psi_{4}-\Psi_{2}^{2})$.

In general the space-time is of type I. However,
in the diagonal solution of the last section there exist
special values for $c$ such that the Weyl tensor is of type
D. These are given by
$c=\pm 1$, $c=\pm\sqrt{1-N^{2}}$  and $c=\pm(1\pm\sqrt{4-N^2})$.

In cosmology one is interested in the evolution of a
space-time with time. Therefore future timelike and future
null infinity will be discussed here.

Following \cite{sta} $Z$ will be taken of the form
\begin{eqnarray}
Z=te^{-2\psi}(1+i2\chi)
\label{Z}
\end{eqnarray}
where $\psi$ and $\chi$ are functions of $t$ and $f$.
In this case $\psi$ and $\chi$ enter more symmetric in the 
Ernst equation and show the same asymptotic behaviour,
which in both cases will be ${\cal O}(t^{-\frac{1}{2}})$ for
$\psi$ and $\chi$. This  asymptotic behaviour show the 
dilaton and axion as well. This is not surprising since
the linear terms in the Ernst equation, the equations for
the axion and the dilaton have the same structure (cf
(\ref{ernst})-(\ref{b_fg})).

At future timelike infinity ($t\rightarrow\infty$, $z$ fixed)
the Weyl scalars behave as

\begin{eqnarray}
\Psi_{0}&\sim& e^{M}{\cal O}(t^{-\frac{1}{2}})\nonumber\\
\Psi_{2}&\sim& e^{M}{\cal O}(t^{-1})\nonumber\\
\Psi_{4}&\sim& e^{M}{\cal O}(t^{-\frac{1}{2}}).
\end{eqnarray}

This shows that the space-time does not approach a Kasner 
model (cf. equations (\ref{wey-kas1})--(\ref{wey-kas3})). 
This can be compared with the future time inifinty of the
Doroshkevich, Zeldovich, Novikov (DZN) \cite{dor}  \cite{clan}
model for which the components of the Weyl tensor are given by
$\Psi_{0}\sim e^{M}t^{-1}$, $\Psi_{2}\sim e^{M}t^{-2}$,
$\Psi_{4}\sim e^{M} t^{-1}$ and the metric is 
$ds^{2}=e^{2t}(dt^{2}-dx^{2})-t^{d+1}dy^{2}
-t^{1-d}dz^{2}$.
So these models show a similar behaviour of the Weyl scalars
at future time infinity. This suggests that the
nondiagonal Einstein-Rosen space-times approach
an anisotropic model similar to the DZN model. However,
they certainly do not approach the Milne model 
for which $\Psi_{0}\sim\Psi_{2}\sim\Psi_{4}\sim{\cal O}(t^{-2})$.
This can be compared with the asymptotic behaviour of diagonal Einstein-Rosen
axion-dilaton cosmologies. In \cite{jk1}  
exact solutions were found which approach a DZN universe containing
axionic, dilatonic and gravitational waves at late times.
A similar behaviour is also found in electromagnetic Gowdy
universes \cite{char}.

To investigate the behaviour of the Weyl tensor at
future null infinity one can use one null coordinate,
say, $f$ and the timelike
coordinate $t$. 

In these variables
the Ernst equation (\ref{ernst})  becomes
\begin{eqnarray}
Z_{tf}+\frac{1}{2}Z_{tt}+\frac{1}{2t}\left(Z_{t}+Z_{f}\right)
=2(Z+\bar{Z})^{-1}\left[Z_{t}Z_{f}+\frac{1}{2}Z_{t}^{2}\right]
\label{ernst1}
\end{eqnarray}

The linear part of the real and imaginary parts of
(\ref{ernst1}) yield to two cylindrical wave equations 
for $\psi$ and $\chi$, respectively. These can be solved by
Bessel functions (cf (\ref{v-sol})). Their asymptotic behaviour
in the limit $t\rightarrow\infty$, $f$ fixed, imply that
both $\psi$ and $\chi$ behave as $a(f)t^{-\frac{1}{2}}$,
if gravitosolitonic contributions are neglected. 
These were discussed for diagonal metrics in the 
last section. They seem to lead to a cosmological model
only in the spatially homogeneous case.

In terms of the coordinates $t$ and $f$ the non-vanishing
components of the Weyl tensor are given by

\begin{eqnarray}
\Psi_{0}&=&\frac{e^{M}}{4}(g')^{2}\left[\frac{\bar{Z}_{tt}}
{Z+\bar{Z}}+\left(\frac{3}{2t}-2t\left[
\frac{Z_{t}\bar{Z}_{t}}{(Z+\bar{Z})^{2}}+
\frac{1}{4}\left(\phi_{t}^{2}+e^{2\phi}b_{t}^{2}\right) \right]\right)
\frac{\bar{Z}_{t}}{Z+\bar{Z}}\right.\nonumber\\
& & \left.-2\frac{\bar{Z_{t}}^{2}}{(Z+\bar{Z})^{2}}\right]
\\
\nonumber\\
\Psi_{2}&=&-\frac{e^{M}}{24}f'g'\left[\frac{3}{2t^{2}}
-12\frac{(Z_{f}+\frac{1}{2}Z_{t})\bar{Z}_{t}}
{(Z+\bar{Z})^{2}}-
(\phi_{f}+\frac{1}{2}\phi_{t})\phi_{t}
-e^{2\phi}(b_{f}+\frac{1}{2}b_{t})b_{t}
\right]
\\
\nonumber\\
\Psi_{4}&=&e^{M}(f')^{2}\left[(Z+\bar{Z})^{-1}
(Z_{ff}+Z_{ft}+\frac{1}{4}Z_{tt})
\right.\nonumber\\
&&
+\left[\frac{3}{4t}-4t\left(\frac{(Z_{f}+\frac{1}{2}Z_{t})
(\bar{Z}_{f}+\frac{1}{2}\bar{Z}_{t})}{(Z+\bar{Z})^{2}}
+\frac{1}{4}\left[
(\phi_{f}+\frac{1}{2}\phi_{t})^{2}
+e^{2\phi}(b_{f}+\frac{1}{2}b_{t})^{2}
\right]
\right)\right]
\frac{Z_{f}+\frac{1}{2}Z_{t}}{Z+\bar{Z}}\nonumber\\
&&
\left.
-2\frac{(Z_{f}+\frac{1}{2}Z_{t})^{2}}{(Z+\bar{Z})^{2}}
\right]
\end{eqnarray}

With $Z$ given by (\ref{Z}) and  the asymptotic behaviour of
$\psi$ and $\chi$ 
the components of the Weyl tensor behave as follows

\begin{eqnarray}
\Psi_{0}&\sim&\frac{e^{M}}{4}(g')^{2}[{\cal O}(t^{-2})]\\
\Psi_{2}&\sim& -\frac{e^{M}}{24}f'g'[{\cal O}(t^{-\frac{3}{2}})]\\
\Psi_{4}&\sim& e^{M}(f')^{2}[{\cal O}(t^{-\frac{1}{2}})]
\end{eqnarray}

So as $t\rightarrow\infty$ the component $\Psi_{4}$ dominates and the
space-time approaches a plane wave space-time at future null
infinity.

This behaviour was already found in the exact non-diagonal solution
of the last section. There the two cases of an ingoing and 
outgoing wave had been stated explicitly. The same applies here 
as well. Interchanging the null coordinates implies an interchange of the 
behaviour of $\Psi_{0}$ and $\Psi_{4}$. 
Hence outgoing and ingoing waves are exchanged.

The equations for dilaton and axion are given by

\begin{eqnarray}
\phi_{ft}+\frac{1}{2}\phi_{tt}+\frac{1}{2t}[\phi_{f}+\phi_{t}]
-e^{2\phi}b_{t}[b_{f}+\frac{1}{2}b_{t}]&=&0\\
b_{ft}+\frac{1}{2}b_{tt}+\frac{1}{2t}[b_{f}+b_{t}]
+b_{t}[\phi_{f}+\frac{1}{2}\phi_{t}]
+\phi_{t}[b_{f}+\frac{1}{2}b_{t}]&=&0.
\end{eqnarray}

Hence, in the limit $t\rightarrow\infty$ the behaviour of the dilaton and
axion is described by,
\begin{eqnarray}
\phi\sim{\cal O}(t^{-\frac{1}{2}})\\
b\sim{\cal O}(t^{-\frac{1}{2}}).
\end{eqnarray}

This leads to a peeling-off behaviour of the components of 
the antisymmetric tensor field strength $H$. 
This is similar to the Einstein-Maxwell case where a 
peeling-off behaviour for the components of the Maxwell
tensor is found in addition to that
of the Weyl tensor \cite{carm1}.

Equation (\ref{H1}) implies that the non-vanishing 
components of the antisymmetric tensor field strength
in the null tetrad are given by

\begin{eqnarray}
H_{(1)(3)(4)}&=&-i\frac{1}{2}e^{2\phi}e^{\frac{M}{2}}f'
(b_{f}+\frac{1}{2}b_{t})\\
H_{(2)(3)(4)}&=&i\frac{1}{2}e^{2\phi}e^{\frac{M}{2}}g'b_{t}.
\end{eqnarray}

This leads to the asymptotic behaviour,
\begin{eqnarray}
H_{(2)(3)(4)}&\sim & e^{2\phi}e^{\frac{M}{2}}{\cal O}(t^{-\frac{3}{2}})\\
H_{(1)(3)(4)}&\sim & e^{2\phi}e^{\frac{M}{2}}{\cal O}(t^{-\frac{1}{2}}).
\end{eqnarray}

\section{Discussion}

The asymptotic behaviour of a large class of inhomogeneous 
axion-dilaton cosmologies has been investigated.
At future time infinity the Weyl scalars approach those
of an anisotropic space-time. At future null infinity
the radiative component of the Weyl scalars dominates.
This might be interpreted as that this class of inhomogeneous 
string cosmologies approaches an 
expanding anisotropic background on which gravitational 
waves propagate. Axion and dilaton become negligible at
late times.

A peeling-off behaviour of the components of the Weyl tensor has been 
found similar to that in the case of vacuum general relativity. 
The components of the antisymmetric tensor field strength
also show a peeling-off behaviour. This is analog to the 
asymptotic behaviour of the Maxwell tensor in the Einstein-Maxwell
case.

The future asymptotic state in general relativity can be interpreted as the
past asymptotic state for the pre-big-bang scenario.
In the class of inhomgeneous metrics considered here, this means that
an anisotropic model and not the Milne universe
is a likely past attractor.

Furthermore, the type of metric which was studied here 
might also be relevant to the recently proposed 
pre-big-bang bubble picture. 
Together with appropriate boundary conditions 
metrics of  the general form (\ref{1}) also describe colliding 
plane wave space-times.
This might be used to find a ``dynamical'' description of the formation of 
pre-big-bang bubbles \cite{buon}.

\vspace{1cm}

{\bf Acknowledgments}
I would like to thank J.D. Barrow, A. Feinstein and G. Veneziano
for discussions. I am especially grateful to A. Feinstein for
his valuable comments on an earlier draft of this manuscript.
Financial support from the Tomalla foundation is acknowledged.

\section*{Appendix: Components of the Weyl and Ricci tensors}

Here all the non-vanishing components of the Weyl tensor
and the Ricci tensor in the Newman-Penrose formalism are given.
The signature of the metric is (+ -- -- --) and Einstein's equations are
given by
\begin{eqnarray}
R_{\mu\nu}-\frac{1}{2}g_{\mu\nu}R=-T_{\mu\nu}.
\label{einst}
\end{eqnarray}

The tetrad metric to raise and lower terad indices
is given by
\begin{eqnarray}
\eta_{(a)(b)}=\eta^{(a)(b)}=\left(
\begin{array}{cccc}
0 & 1 & 0 & 0\\
1 & 0 & 0 & 0\\
0 & 0 & 0 &-1\\
0 & 0 & -1 & 0
\end{array}
\right).
\end{eqnarray}
 
Tetrad indices are enclosed in brackets, which will be omitted if
it is unambiguous. Tetrad indices run from 1 to 4.
 
For a space-time with metric (\ref{1}) a null tetrad basis is
provided by \cite{grif}

\begin{eqnarray}
e_{(1)}=e^{(2)}=l_{\mu}=e^{-\frac{M}{2}}
\left\{
\begin{array}{cccc}
1 & 0 & 0 & 0
\end{array}   
\right\}\nonumber\\
e_{(2)}=e^{(1)}=n_{\mu}=e^{-\frac{M}{2}}
\left\{
\begin{array}{cccc}
0 & 1 & 0 & 0
\end{array}
\right\}\nonumber\\
-e_{(3)}=e^{(4)}=-m_{\mu}=e^{-\frac{U}{2}}(Z+\bar{Z})^{-\frac{1}{2}}
\left\{
\begin{array}{cccc}
0 & 0 & 1 & -i\bar{Z}
\end{array}
\right\}\nonumber\\
-e_{(4)}=e^{(3)}=-\bar{m}_{\mu}=e^{-\frac{U}{2}}(Z+\bar{Z})^{-\frac{1}{2}}
\left\{
\begin{array}{cccc}
0 & 0 & 1 & iZ
\end{array}
\right\}
\end{eqnarray}
 
and the corresponding contravariant components are given by
 
\begin{eqnarray}   
l^{\mu}=e^{\frac{M}{2}}
\left\{
\begin{array}{cccc}
0 & 1 & 0 & 0
\end{array}
\right\}\nonumber\\
n^{\mu}=e^{\frac{M}{2}}
\left\{
\begin{array}{cccc}
1 & 0 & 0 & 0
\end{array}
\right\}\nonumber\\
m^{\mu}=e^{\frac{U}{2}}(Z+\bar{Z})^{-\frac{1}{2}}
\left\{
\begin{array}{cccc}
0 & 0 & \bar{Z} & -i
\end{array}
\right\}\nonumber\\
\bar{m}^{\mu}=e^{\frac{U}{2}}(Z+\bar{Z})^{-\frac{1}{2}}
\left\{
\begin{array}{cccc}
0 & 0 & Z & i 
\end{array}
\right\}.
\end{eqnarray}
 
where $(x^{0},x^{1},x^{2},x^{3})\equiv(u,v,x,y)$.
 
The Newman-Penrose spin coefficients have been calculated
using Maple \cite{maple},
 
\begin{eqnarray}   
\lambda&=&-e^{\frac{M}{2}}\frac{Z_{u}}{Z+\bar{Z}}\nonumber\\
\mu&=&-\frac{1}{2}e^{\frac{M}{2}}U_{u}\nonumber\\
\rho&=&\frac{1}{2}e^{\frac{M}{2}}U_{v}\nonumber\\
\sigma&=&e^{\frac{M}{2}}\frac{\bar{Z}_{v}}{Z+\bar{Z}}\nonumber\\
\gamma&=&\frac{1}{4}e^{\frac{M}{2}}\left[M_{u}-\frac{Z_{u}-\bar{Z}_{u}}
{Z+\bar{Z}}\right]\nonumber\\
\epsilon&=&-\frac{1}{4}e^{\frac{M}{2}}\left[M_{v}+  
\frac{Z_{v}-\bar{Z}_{v}}{Z+\bar{Z}}\right]\nonumber\\
\nu&=&\pi=\tau=\kappa=\alpha=\beta=0
\end{eqnarray}
 
$F_{u}$ denotes the partial derivative $\frac{\partial F}{\partial u}$.
 
In the Newman-Penrose formalism the components of the Weyl tensor 
are given by five complex scalars \cite{chandra}
\begin{eqnarray}   
\Psi_{0}&=&-C_{(1)(3)(1)(3)}=-C_{\mu\nu\lambda\kappa}l^{\mu}m^{\nu}
l^{\lambda}m^{\kappa}\nonumber\\
\Psi_{1}&=&-C_{(1)(2)(1)(3)}=-C_{\mu\nu\lambda\kappa}l^{\mu}n^{\nu}
l^{\lambda}m^{\kappa}\nonumber\\
\Psi_{2}&=&-C_{(1)(3)(4)(2)}=-C_{\mu\nu\lambda\kappa}l^{\mu}m^{\nu}
\bar{m}^{\lambda}n^{\kappa}\nonumber\\
\Psi_{3}&=&-C_{(1)(2)(4)(2)}=-C_{\mu\nu\lambda\kappa}l^{\mu}n^{\nu}
\bar{m}^{\lambda}n^{\kappa}\nonumber\\
\Psi_{4}&=&-C_{(2)(4)(2)(4)}=-C_{\mu\nu\lambda\kappa}n^{\mu}\bar{m}^{\nu}
n^{\lambda}\bar{m}^{\kappa}
\end{eqnarray}
 
and the components of the Ricci tensor are denoted by the following
four real and three complex scalars \cite{chandra}
 
\begin{eqnarray}
\Phi_{00}=-\frac{1}{2}R_{(1)(1)} & &
\Phi_{22}=-\frac{1}{2}R_{(2)(2)}\nonumber\\
\Phi_{02}=-\frac{1}{2}R_{(3)(3)} & &
\Phi_{20}=-\frac{1}{2}R_{(4)(4)}\nonumber\\
\Phi_{11}=-\frac{1}{4}(R_{(1)(2)}+R_{(3)(4)})
& & \Phi_{01}=-\frac{1}{2}R_{(1)(3)}\nonumber\\
\Lambda=\frac{1}{24}R=\frac{1}{12}(R_{(1)(2)}-R_{(3)(4)})& &
\Phi_{10}=-\frac{1}{2}R_{(1)(4)}\nonumber\\
\Phi_{12}=-\frac{1}{2}R_{(2)(3)} & &
\Phi_{21}=-\frac{1}{2}R_{(2)(4)}\nonumber
\end{eqnarray}

Using the relations given in \cite{chandra}  the components of the
Weyl and the Ricci tensor are derived.

The components of the Weyl tensor are given by
 
\begin{eqnarray}
\Psi_{0}&=&\frac{e^{M}}{(Z+\bar{Z})^{2}}
\left[(Z+\bar{Z})(\bar{Z}_{vv}+M_{v}\bar{Z}_{v}
-U_{v}\bar{Z}_{v})-2(\bar{Z}_{v})^{2}\right]\nonumber\\
\Psi_{2}&=&-\frac{1}{4}e^{M}\left(2U_{uv}-U_{u}U_{v}-
4\frac{Z_{u}\bar{Z}_{v}}{(Z+\bar{Z})^{2}}\right)
-2\Lambda\nonumber\\
\Psi_{4}&=&\frac{e^{M}}{(Z+\bar{Z})^{2}}
\left[(Z+\bar{Z})(Z_{uu}+M_{u}Z_{u}-U_{u}Z_{u})-2(Z_{u})^{2}
\right]\nonumber\\
\Psi_{1}&=&\Psi_{3}=0
\end{eqnarray}  
 
and the non-vanishing components of the Ricci tensor are 
 
\begin{eqnarray}
\Phi_{00}&=&\frac{e^{M}}{4}\left[2U_{vv}-(U_{v})^{2}+
2M_{v}U_{v}-4\frac{Z_{v}\bar{Z}_{v}}{(Z+\bar{Z})^{2}}
\right]\nonumber\\
\Phi_{02}&=&-\frac{1}{2}\frac{e^{M}}{(Z+\bar{Z})^{2}}
\left[(Z+\bar{Z})(2\bar{Z}_{uv}-U_{u}\bar{Z}_{v}-U_{v}\bar{Z}_{u})
-4\bar{Z}_{u}\bar{Z}_{v}\right]\nonumber\\
\Phi_{11}&=&\frac{e^{M}}{8}\left[2M_{uv}+U_{u}U_{v}
-2\frac{Z_{u}\bar{Z}_{v}+\bar{Z}_{u}Z_{v}}{(Z+\bar{Z})^{2}}
\right]
\nonumber\\
\Phi_{22}&=&\frac{e^{M}}{4}\left[2U_{uu}-(U_{u})^{2}
+2U_{u}M_{u}-4\frac{Z_{u}\bar{Z}_{u}}{(Z+\bar{Z})^{2}}
\right]\nonumber\\
\Lambda&=&-\frac{e^{M}}{24}\left[
2M_{uv}+4U_{uv}-3U_{u}U_{v}-2\frac{Z_{u}\bar{Z}_{v}+\bar{Z}_{u}Z_{v}}
{(Z+\bar{Z})^{2}}\right].
\end{eqnarray}

The non-vanishing components and the trace 
in a space-time with metric (\ref{1})
are given by
\begin{eqnarray}
T_{00}=\frac{1}{2}[\phi_{u}^{2}+e^{2\phi}b_{u}^{2}] & &
T_{11}=\frac{1}{2}[\phi_{v}^{2}+e^{2\phi}b_{v}^{2}]\nonumber\\
T_{22}=\frac{e^{M-U}}{Z+\bar{Z}}[\phi_{u}\phi_{v}
+e^{2\phi}b_{u}b_{v}]& &
T_{23}=\frac{i}{2}e^{M-U}\frac{Z-\bar{Z}}{Z+\bar{Z}}[\phi_{u}\phi_{v}
+e^{2\phi}b_{u}b_{v}]\nonumber\\
T_{33}=e^{M-U}\frac{Z\bar{Z}}{Z+\bar{Z}}[\phi_{u}\phi_{v}
+e^{2\phi}b_{u}b_{v}]
& &
T=-e^{M}[\phi_{u}\phi_{v}+e^{2\phi}b_{u}b_{v}]
\end{eqnarray}

and  using Einstein's equations  the only
non-vanishing
components of the Ricci tensor are as follows

\begin{eqnarray}
R_{00}=-\frac{1}{2}[\phi_{u}^{2}+e^{2\phi}b_{u}^{2}]\;\;\;\; &\;\;\;\;
R_{01}=-\frac{1}{2}[\phi_{u}\phi_{v}+e^{2\phi}b_{u}b_{v}]\;\;\;\;
&\;\;\;\;
R_{11}=-\frac{1}{2}[\phi_{v}^{2}+e^{2\phi}b_{v}^{2}].
\end{eqnarray}

So the non-vanishing
Newman-Penrose scalars $\Phi_{ab}$ and $\Lambda$ are given 
by
\begin{eqnarray}
\Phi_{00}=\frac{e^{M}}{4}[\phi_{v}^{2}+e^{2\phi}b_{v}^{2}]\;\;\;\;&\;\;\;\;
\Phi_{11}=\frac{e^{M}}{8}[\phi_{u}\phi_{v}+e^{2\phi}b_{u}b_{v}]\nonumber\\
\Phi_{22}=\frac{e^{M}}{4}[\phi_{u}^{2}+e^{2\phi}b_{u}^{2}]\;\;\;\;&\;\;\;\;
\Lambda=-\frac{e^{M}}{24}[\phi_{u}\phi_{v}+e^{2\phi}b_{u}b_{v}].
\end{eqnarray}


\begin{thebibliography}{}
\bibitem{gasp} M. Gasperini, G. Veneziano, Astropart. Phys.
{\bf 1}, 317 (1993); G. Veneziano, ``A simple/short introduction
to pre-big-bang physics/cosmology'', hep-th/9802057;
for an updated collection of papers on string cosmology, see
{\rm http://www.to.infn.it/ $\sim$gasperin}.

\bibitem{buon} A. Buonanno, T. Damour, G. Veneziano,
Nucl. Phys. B {\bf 543}, 275 (1999).

\bibitem{clan} D. Clancy, J.E. Lidsey, R. Tavakol,
Phys. Rev. D {\bf 59}, 063511 (1999).

\bibitem{Buon2} A. Buonanno, K.A. Meissner, C. Ungarelli, 
G. Veneziano, Phys. Rev. D {\bf 57}, 2543 (1998). 

\bibitem{jk1}
J.D. Barrow, K.E. Kunze, Phys. Rev. D {\bf 56}, 741 (1997).

\bibitem{p3}
D. Clancy, A. Feinstein, J.E. Lidsey, R. Tavakol, ``Inhomogeneous
Einstein-Rosen String Cosmology'', gr-qc/9901061;
A. Feinstein, R. Lazkoz, M.A. Vazquez-Mozo, Phys. Rev. D{\bf 56},
5166 (1997).

\bibitem{sachs} H. Bondi, M.G.J. van der Burg, A.W.K. Metzner,
Proc. Roy. Soc. (London) {\bf A 269}, 21 (1962);
R.K. Sachs, Proc. Roy. Soc. (London) {\bf 270}, 103 (1962);
E.T. Newman, R. Penrose, J. Math. Phys. {\bf 3}, 566 (1962);
R.K. Sachs in {\sl Relativity, Groups and Topology},
ed. by C. DeWitt, B. DeWitt, Gordon and Breach, Science
Publishers, Inc. New York (1964). 

\bibitem{carm}
M. Carmeli, A. Feinstein, Phys. Lett. B {\bf 103A}, 318 (1984);
Int. J. Theor. Phys. {\bf 24}, 1009 (1985).


\bibitem{sta}
J.J. Stachel, Journ. of Math. Physics {\bf 7}, 1321 (1966).

\bibitem{tom}
K. Tomita, Prog. Theor. Phys. {\bf 59}, 1150 (1978).


\bibitem{r4}
E.J. Copeland, A. Lahiri, D. Wands, Phys. Rev. D, {\bf 50},
4868 (1994).


\bibitem{grif}
J.B. Griffiths, {\it Colliding Plane Waves in General
Relativity}, Oxford University Press, Oxford, UK, (1991).


\bibitem{chandra}
S. Chandrasekhar, {\it The Mathematical Theory of
Black Holes}, Oxford University Press, Oxford, UK, (1992).


\bibitem{bata}
N.A. Batakis, Class. Quant. Grav. {\bf 13}, L95 (1996).


\bibitem{mirror}
I. Bakas, Nucl. Phys. B {\bf 428}, 374 (1994);
J.E. Lidsey, Class. Quant. Grav. {\bf 15}, L77 (1998).

\bibitem{wain} J. Wainwright, W.C.W. Ince,
B.J. Marshman, Gen. Rel. Grav. {\bf 10}, 259 (1979).

\bibitem{kitch} D.W. Kitchingham, Class. Quant. Grav. {\bf 1},
677 (1984).

\bibitem{verd}
E. Verdaguer, Phys. Rep. {\bf 229,} 1 (1993).


\bibitem{ellis} G.F.R. Ellis, M.A.H. MacCallum, Comm. Math. Phys. {\bf 12},
108 (1969).

\bibitem{carm2} M. Carmeli, Ch. Charach, S. Malin, Phys. Rep. {\bf 76},
79 (1981).

\bibitem{hew} C.G. Hewitt, J. Wainwright, Class. Quant. Grav. {\bf 10},
99 (1993).

\bibitem{dinv}
R.A. D'Inverno, R.A. Russell-Clark, Journ. of Math. Physics
{\bf 12}, 1258 (1971).


\bibitem{dor} A.G. Doroshkevich, Y.B. Zeldovich, I.D. Novikov, Sov. Phys.
JETP {\bf 26}, 408 (1968).

\bibitem{char} Ch. Charach, Phys. Rev. D {\bf 19}, 3516 (1979);
Ch. Charach, S. Malin, Phys. Rev. D {\bf 21}, 3284 (1980).

\bibitem{carm1} J.N. Goldberg, R.P. Kerr, Journ. of Math. Phys. {\bf 5},
172 (1964);
 M. Carmeli, ``Group Theory and General Relativity'',
McGraw-Hill, (1977).

\bibitem{maple}
Maple V.3   Waterloo Maple Software (1994).

\end{thebibliography}
\end{document}